\documentclass[lettersize,journal]{IEEEtran}

\usepackage{acronym}
\usepackage{amsfonts}
\usepackage{amsmath}
\usepackage{amsthm}
\usepackage{balance}  
\usepackage{booktabs}  
\usepackage{cite}
\usepackage[inline,shortlabels]{enumitem}
\usepackage{graphicx}
\usepackage{hyperref}
\usepackage[switch]{lineno}  
\usepackage{subcaption}  
\usepackage[textwidth=1.2cm,textsize=footnotesize]{todonotes}
\usepackage{xcolor}
\usepackage{textcomp}
\usepackage{comment}

\usepackage{commands}

\usepackage{uppaal}

\graphicspath{{images/}}

\newif\ifanonymous
\anonymousfalse

\begin{document}

\title{%
Automated Detection and Mitigation of Dependability Failures in Healthcare Scenarios through Digital Twins
}

\ifanonymous{
\author{
\IEEEauthorblockN{
Anonymous Author(s)
}}
\else{
\author{
\IEEEauthorblockN{
Bruno Guindani\IEEEauthorrefmark{1}, 
Matteo Camilli\IEEEauthorrefmark{1}, 
Livia Lestingi\IEEEauthorrefmark{1}, 
Marcello M. Bersani\IEEEauthorrefmark{1}}%

\IEEEauthorblockA{
\IEEEauthorrefmark{1}Dipartimento di Elettronica, Informazione e Bioingegneria\\
Politecnico di Milano, Via Ponzio 34/5, 20133 Milan, Italy\\
Email: \{name\}.\{surname\}@polimi.it}}
}
\fi

\maketitle

\begin{abstract}
Medical \acp{CPS} integrating Patients, Devices, and healthcare personnel (Physicians) form safety-critical \acs{PDP} triads whose dependability is challenged by system heterogeneity and uncertainty in human and physiological behavior.
While existing clinical decision support systems support clinical practice, there remains a need for proactive, reliability-oriented methodologies capable of identifying and mitigating failure scenarios before patient safety is compromised.
This paper presents \nameExt{}, a methodology based on a closed-loop \ac{DT} paradigm for dependability assurance of medical \acp{CPS}.
The approach combines \acl{SHA} modeling, data-driven learning of patient dynamics, and \acl{SMC} with an offline critical scenario detection phase that integrates model-space exploration and diversity analysis to systematically identify and classify scenarios violating expert-defined dependability requirements.
\nameExt{} also supports the automated synthesis of mitigation strategies, enabling runtime feedback and control within the \ac{DT} loop.
We evaluate \nameExt{} on a representative use case study involving a pulmonary ventilator.
Results show that, in 87.5\% of the evaluated scenarios, strategies synthesized through formal game-theoretic analysis stabilize patient vital metrics at least as effectively as human decision-making, while maintaining relevant metrics 20\% closer to nominal healthy values on average.
\end{abstract}

\acresetall

\section{Introduction}
\label{sec:intro}

Under hospitalized care, patients often interact with medical devices under the supervision of healthcare personnel.
This \ac{PDP} triad constitutes a medical \ac{CPS} requiring strict operational safety and specialized design~\cite{dhaked2024mcps}.

Medical practice defines the conditions under which a patient is considered clinically stable based on their vital parameters.
We refer to situations in which the \ac{PDP} triad cannot maintain such clinical stability as \emph{failure scenarios}, whereas the \ac{PDP} is said to be \emph{dependable} when coordinated practitioner and device actions successfully preserve patient stability.
Failure scenarios are commonly recognized and addressed by relying on human expertise and \acp{CDSS}.
\ac{CDSS} software~\cite{sutton2020cdss} supports clinicians by providing patient-specific knowledge and information to enhance care and reduce errors at multiple stages of the clinical process.
Engineering automated monitoring and control mechanisms for such medical \acp{CPS} remains challenging due to the inherent complexity of their interacting components and the uncertainty of human decision-making, which is influenced by individual expertise and environmental conditions.
The US Institute of Medicine emphasizes that safe healthcare systems should ``adopt a proactive approach that allows for examining processes of care for threats to safety'' and incorporate mechanisms for ``learning from errors''~\cite{iom2000toerr}.

In this paper, we present an extension of \name{}\footnote{\namefull{}.}~\cite{guindani2025detecting}, a methodology for the proactive identification of dependability failures in \ac{PDP} \acp{CPS}.
The approach adopts the \ac{DS} paradigm and combines automata-based modeling, data-driven learning of patient dynamics, and formal verification with model space exploration techniques to systematically uncover failure scenarios, where the \ac{PDP} triad violates given \emph{dependability requirements}.

Our extended version, referred to as \nameExt{}, complies with the \ac{DT} paradigm~\cite{abanda2025industry}.
As such, it enables a bidirectional digital representation of a physical system, where monitored data continuously updates the digital model, which can, in turn, influence the physical system through feedback and control.
Beyond retrospective analysis, diagnostics, and monitoring, a \ac{DT} enables simulation, prediction, and decision support by closing the loop between physical and digital domains.
In addition to the identification of failure scenarios, our extension \nameExt{} supports:
\begin{enumerate*}[label=\textit{\roman*)}]
    \item diversity analysis of failures by isolating distinct critical contingencies;
    \item automated synthesis of response strategies to such critical contingencies;
    \item alignment of the \ac{PDP} model with data from the physical system (\emph{feedback} action of the \ac{DT} paradigm) upon failure detection; and
    \item clinical decision support through the recommendation of optimal mitigation actions for ongoing failures (\emph{control} action of the \ac{DT} paradigm).
\end{enumerate*}

The \ac{PDP} model at the core of our methodology captures the functions of medical equipment (including monitors for critical clinical metrics), the practitioner's actions, the patient's behavior, and all relevant interactions between the three entities by adopting \acf{SHA}~\cite{david2011statistical}.
Specifically, we use \emph{hybrid} features to capture non-trivial dynamics of physiological quantities and \emph{stochastic} features to approximate the sources of uncertainty due to human behavior (\eg{} alternative practitioner decisions and delays or sudden changes in patient health).
The \ac{SHA} model is developed by relying on expert knowledge and field-collected data.
For aspects subject to uncertainty (\eg{} sudden changes in patient health), \nameExt{} uses the data-driven automata learning technique \lsha~\cite{lestingi2022model, lestingi2024data} to infer the patient model from real physiological logs.

The resulting parametric model represents a family of \ac{PDP} triads and identifies a design space.
A point in the design space (\ie{} a value-assignment to the triad parameters) corresponds to a specific clinical scenario (\eg{} the onset of an asthma attack).
Our methodology explores the design space through fuzzing~\cite{manes2021art}.
Starting from a reference model based on domain knowledge, multiple variants (mutants) are automatically generated to detect violations of dependability requirements (defined by domain experts).
Specifically, each generated \ac{SHA} mutant is analyzed through \ac{SMC} to calculate the probability that a requirement holds~\cite{legay2010statistical}.
After filtering out unrealistic or clinically irrelevant scenarios, we analyze the diversity of the failures to isolate distinct root causes.
\ac{SHA} scenarios are converted to 1\textonehalf-player-games to synthesize near-optimal strategies through \uppaal{} \stratego{}~\cite{david2015uppaal}, informing physicians on what actions to take to maximize the likelihood of satisfying the requirements.

We empirically evaluate the proposed methodology on a representative \ac{PDP} exemplar involving a \emph{patient} affected by lung disease, a pulmonary ventilator (the \emph{device}), and an intensivist (the \emph{physician}).
Domain knowledge elicited through expert interviews informs both the physician–device model and the specification of dependability requirements, while patient data are generated using a high-fidelity physiological simulator~\cite{bray2019pulse}.
The experimental campaign includes $10$ hours of cumulative acquisition time, including both interviews and simulation sessions.
Results show that, in $87.5\%$ of the evaluated scenarios, the simulated physician guided by the synthesized strategies stabilizes patient vital metrics at least as effectively as the human physician, while maintaining the relevant metrics $20\%$ closer to their nominal healthy values.

In summary, compared to our prior work~\cite{guindani2025detecting}, the main contributions are:
\begin{itemize}
\item an online clinical decision support phase upgrading the previous \ac{DS} to a fully fledged \ac{DT} infrastructure with real-time state alignment and action selection;
\item a diversity analysis of failure scenarios to isolate different root causes;
\item an automated synthesis of control strategies that support medical personnel in selecting actions that ensure the dependability of the \ac{PDP} system; and
\item an extended empirical evaluation covering the newly added elements of the methodology.
\end{itemize}

The paper is structured as follows: \sref{background} reviews preliminary concepts; \sref{application} introduces the \ac{PDP} exemplar serving as running example; \sref{methodology} details the developed methodology; \sref{exp} presents experimental results; \sref{discussion} discusses findings; \sref{rw} surveys related work; and \sref{concl} concludes.

\section{Preliminaries}
\label{sec:background}

\subsection{Formal Analysis}
\label{sec:sha}

\subsubsection{Stochastic Hybrid Automata}
\ac{HA} extend finite-state automata by including real-valued variables that evolve continuously over time according to flow conditions specified by \acp{ODE}~\cite{alur1995algorithmic}, thus modeling systems with complex, non-linear dynamics.
An \ac{HA} is a tuple $\mathcal{H} = (\Sigma, L, X, F, E, I, P)$ where:
\begin{enumerate*}[\textit{\roman*})]
\item $\Sigma$ is a set of actions, partitioned into inputs $(\Sigma_\mathrm{i})$ and outputs $(\Sigma_\mathrm{o})$;
\item $L$ is a set of locations, each representing a distinct operational mode of an agent or a physical phenomenon;
\item $X$ is a set of real-valued variables governing system dynamics, and $G(X)$ is the set of constraints on $X$ (e.g., ${x<10}$);
\item $F = { f_\ell : \ell \in L }$ defines flow conditions, with each $f_\ell$ describing the time evolution of variables in $X$ at location $\ell$ through a system of \acp{ODE} $\dot{X} = f_\ell(X)$;
\item $E \subseteq L \times G(X) \times \Sigma \times 2^X \times L$ is a finite set of edges between locations, each including a guard from $G(X)$, an event from $\Sigma$, and an assignment to variables of $X$ called \emph{resets} (e.g., ${x=0}$);
\item ${I: L \to G(X)}$ assigns \emph{invariants}, i.e., constraints on $X$, to locations;
\item ${P: L \to \mathbb{R}^+}$ assigns exit rates to locations.
\end{enumerate*}
\acf{SHA} extend \ac{HA} with stochastic elements~\cite{david2011statistical} that represent uncertainties in physical phenomena.
For example, if parameter $\theta$ in a flow condition is treated as a random variable s.t. $\dot{X} = f(X, \theta)$ with $\theta \sim \mathcal{D}$, then $X$ follows a stochastic process.

An \ac{SHA} network consists of multiple \acp{SHA} that synchronize on actions in $\Sigma$, ensuring that all the involved automata take an edge simultaneously when an event occurs.
Specifically, an \ac{SHA} can trigger synchronization via an edge labeled ${\sigma \in \Sigma_\mathrm{o}}$, while others react with ${\sigma \in \Sigma_\mathrm{i}}$ at the same time.

\subsubsection{\texorpdfstring{\lsha{}}~ for SHA learning}
Automata learning algorithms largely adopt the teacher-learner structure from \lstar~\cite{angluin1987learning}.
Our approach uses \lsha{}, extending \lstar{} to \ac{SHA}~\cite{lestingi2022model, lestingi2024data}, where the 
training dataset consists of \emph{signals}, \ie{} sequences of timestamp-value pairs $(t_0,s_0), (t_1,s_1), \dots$, with $t_i,s_i \in \mathbb{R}$ and $0\leq t_i \leq t_{i+1}$ for adjacent pairs.

Besides the training dataset, \lsha{} requires:
\begin{enumerate*}[\textit{\roman*})]
\item a set $X_\mathrm{F}$ of real-valued variables whose time dynamics are inferred and constrained by flow conditions;
\item a set $\Sigma$ of actions corresponding to real-world events, whose relationship with changes in the dynamics of variables in $X_\mathrm{F}$ is learned from data;
\item a set $X_\mathrm{M}$ of real-valued variables used to extract events from the collected data.
\end{enumerate*}
Each variable in $X_\mathrm{F}$ and $X_\mathrm{M}$ is associated with a signal in the dataset.
A user-defined \emph{labeling} partial function $\mathcal{L}_{X_\mathrm{M}}$ maps signals to traces (\ie sequences of actions).
Specifically, ${\mathcal{L}_{X_\mathrm{M}}(t) = \sigma \in \Sigma}$ indicates the action corresponding to the event at timestamp $t$.
\lsha{} learns: locations $L$, edges $E$, flow conditions $F$ for variables in $X_\mathrm{F}$, and distributions $\mathcal{D}$ for random parameters in the \acp{ODE}.
The resulting \ac{SHA} is suitable for formal verification and can serve as a predictor for variables in $X_\mathrm{F}$, given the actions in $\Sigma$.

\subsubsection{Strategy Synthesis}
Strategy synthesis for \ac{SHA} networks aims to derive an optimal \emph{controller} policy that resolves nondeterminism to satisfy a quantitative reachability objective under stochastic uncertainty.

An \ac{SHA} network is interpreted as a 1\textonehalf-player timed game~\cite{david2014time}. 
In contrast to classical 2-player games, the environment is stochastic rather than adversarial, and reachability objectives are therefore addressed via stochastic strategies over controllable actions.
Accordingly, the output action alphabet is partitioned as \emph{controllable} and \emph{uncontrollable} ${\Sigma_\mathrm{o}=\Sigma_\mathrm{o,c}\cup\Sigma_\mathrm{o,u}}$.
A \emph{stochastic strategy} is a density function $\optstrat: L \times \Real_+^X \times\Real_+ \times \Sigma_\mathrm{o, c} \to \Real_+$, defining the likelihood of selecting a specific controllable action $\sigma \in \Sigma_\mathrm{o,c}$ after a delay $d \in \Real_+$ for any possible system configuration $(l,\nu) \in L \times \Real_+^X$.
Strategy synthesis reduces to a \emph{cost-bounded reachability problem} that, given a bound $b \in \Real_+^X$, entails the computation of a \emph{stochastic strategy} guaranteeing the network reaches a \emph{safe state}, \ie a location in ${G \subseteq L}$ such that the accumulated costs are bounded by $b$.
When dealing with stochastic strategies in timed games, it is common practice to restrict the scope to so-called \emph{non-lazy} strategies~\cite{DavidLLMP13, david2014time}.
Informally, these strategies are designed to avoid ``procrastination'', as they ensure that controllable discrete actions are not unnecessarily delayed when they contribute to reaching the goal. 
Instead of waiting for an arbitrary amount of time, a non-lazy strategy decides whether to act immediately or to wait for an environmental action, at any time instant.
Non-lazy strategies can be effectively recast as deterministic strategies. 
For each configuration $(l,v)$, one can simply select the most promising action. 
The deterministic choice maximizes the probability of success, effectively collapsing the stochastic distribution into a single, optimal decision.
More formally, the resulting strategy derived from $\optstrat$ is a mapping $\controller: L \times \Real_+^X \to \Sigma_\mathrm{o,c}$.
\uppaal{} \stratego{}~\cite{david2015uppaal} is adopted in this work to model the reachability game and compute the stochastic strategy.

\subsection{Fuzzing}
\label{subsec:fuzzing}
Fuzzing~\cite{manes2021art} is an automated testing approach that systematically generates inputs.
Instead of relying on manually crafted test cases, fuzzing explores a wide range of program behaviors to uncover defects, often using runtime feedback (\eg code coverage) to guide the process.
Inputs can be generated from scratch or derived from existing examples to reveal unexpected behavior, such as assertion violations.

Mutational fuzzing~\cite{bohme2016CCS} produces new test cases by applying transformations to a set of valid inputs (seeds).
This approach assumes that small modifications to valid inputs can trigger new behaviors while largely preserving validity.
Feedback mechanisms, such as code coverage or estimated likelihood of failure, are used to prioritize mutations, enabling the fuzzer to incrementally explore untested behavior.

Search-based fuzzing (or testing)~\cite{bohme2017CCS} extends mutational fuzzing by applying optimization strategies to guide input generation toward specific objectives.
In this setting, test case generation is treated as an optimization problem, where candidate inputs are evaluated with a fitness function often based on execution depth, branch coverage, or proximity to requirements violations.
Techniques like evolutionary strategies or genetic algorithms~\cite{deb2002fast} allow the fuzzer to focus on input regions with higher potential to reveal failures.

\section{Running Example: Respiratory Intensive Care} 
\label{sec:application}

\begin{figure}[t]
    \centering
    \includegraphics[width=\columnwidth]{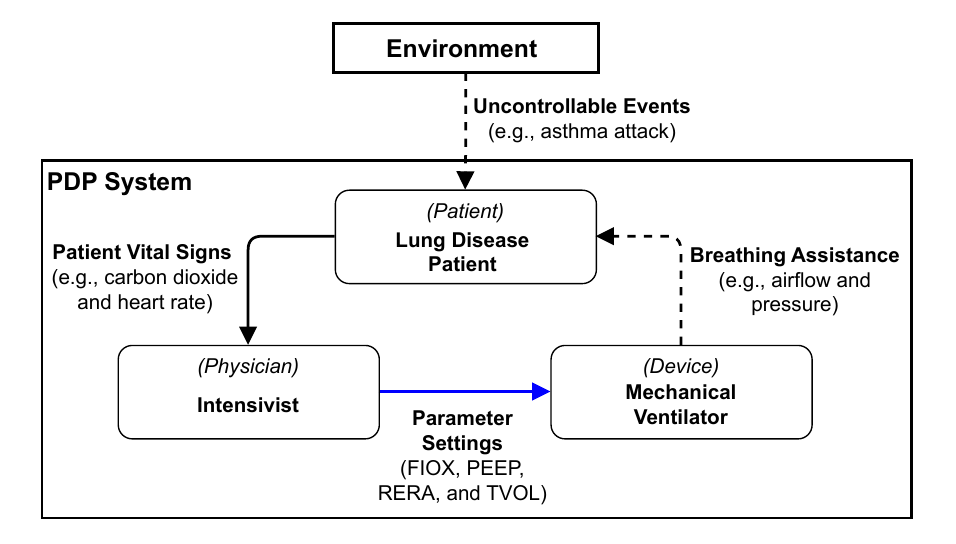}
    \caption{\ac{PDP} triad used as a running example showing interactions between the agents and the environment: an arrow from agent $a$ to $b$ represents an event that $a$ triggers directed to $b$. Controllable events are in blue, uncontrollable but observable events are in solid black, while uncontrollable and not directly observable events are represented by dashed arrows.}
    \label{fig:pdp}
\end{figure}

Our scenarios of interest occur in the hospital \ac{ICU} and involve a patient with lung disease admitted to the \ac{ICU}, the medical staff responsible for their care, and the mechanical ventilator (the device) operated by the staff to provide life support (see \fref{pdp}).

The intensivist can affect the patient’s respiratory state~\cite{owens2018ventilator} by toggling the ventilator’s operation and adjusting the four parameters\footnote{$\param{TVOL}$ values depend on patient characteristics such as body mass~\cite{owens2018ventilator}. The reported range corresponds to values used in our experiments.} in \tref{ventilator}~\cite{bersten2013oh}. 
\begin{table}[tb]
\caption{Ventilator parameters.}
\label{tab:ventilator}
    \centering
    \begin{tabular}{cccc}
        \toprule
        acronym & full name & range & unit \\
        \midrule
        $\param{FIOX}$ & fraction of inspired oxygen      &     $[0, 1]$ &   N/A \\
        $\param{PEEP}$ & positive end-expiratory pressure &    $[5, 25]$ & cmH2O \\
        $\param{RERA}$ & forced respiration rate          &    $[6, 18]$ & $\text{min}^{-1}$ \\
        $\param{TVOL}$ & forced tidal volume              & $[300, 500]$ & mL \\
        \bottomrule
    \end{tabular}
\end{table}
Actions are applied to patients according to standard procedures~\cite{owens2018ventilator}, manipulating ventilator parameters to mitigate adverse events.
Appropriate interventions are determined by the intensivist through continuous monitoring of patient's key vital signs---carbon dioxide $\param{CD}$, heart rate $\param{HR}$, oxygen saturation $\param{OS}$, respiration rate $\param{RR}$, and tidal volume $\param{TV}$--- with actions based on changes that require medical attention.
As illustrated in \fref{pdp}, the intensivist has full control over the ventilator but cannot directly observe events affecting the patient’s health, inferring conditions such as an asthma attack through observed variations in vital signs.

This \ac{PDP} triad forms a complex \ac{CPS}, where human agents interact with cyber-physical devices and exercise judgment in operating them.
Uncertainty arises from both human behavior (the physician’s decisions) and environmental events that influence the patient’s vitals (\eg asthma attacks).
These factors highlight the importance of \ac{DT} paradigms capable of accurately modeling all agents in the \ac{PDP} triad.
Moreover, detecting and analyzing critical scenarios that reflect reductions in practitioner effectiveness through the \ac{DT} can aid medical staff in preventing patient-threatening situations.

\section{Methodology}
\label{sec:methodology}

\begin{figure*}[t]
    \centering
    \includegraphics[width=2\columnwidth]{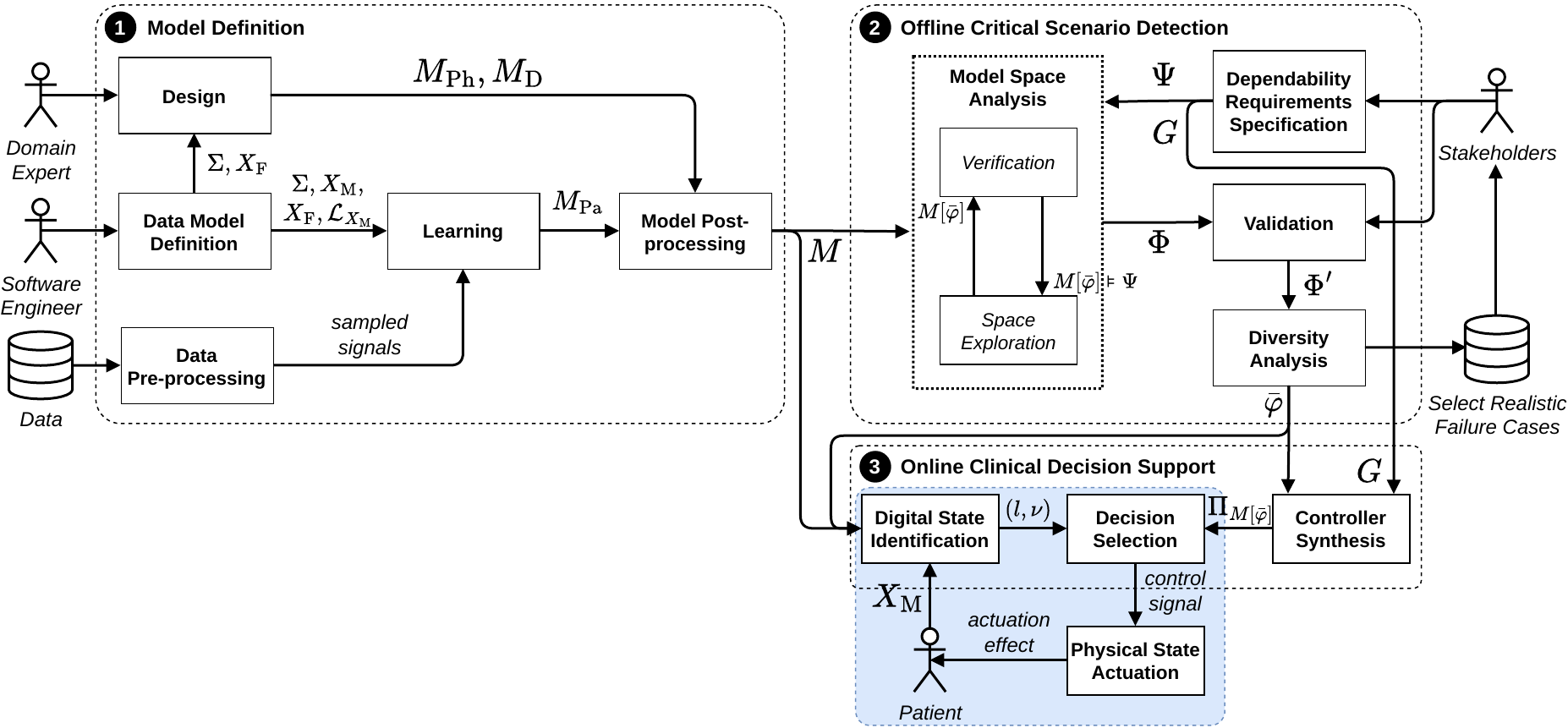}
    \caption{Overview of proposed methodology, highlighting the three \emph{phases} (dashed boxes), the data flow between phase \emph{activities} (solid boxes), and the actors involved in activities requiring human input. The \ac{DT} alignment loop is in blue.}
    \label{fig:overview}
\end{figure*}

\nameExt{} identifies potentially critical scenarios that may endanger the patient's health status (the \emph{failures}), and synthesizes action strategies to provide the physician with clinical decision support.
As depicted in \fref{overview}, the methodology includes three phases and requires the collaboration of actors with a medical background (\ie the \emph{domain experts}) and technical expertise (\ie \emph{software engineers}).
In particular, it generates models of \ac{PDP} triads by leveraging domain knowledge and data collected in real-world settings (the \textbf{Model Definition} phase).
Model are then analyzed to determine whether they satisfy set dependability requirements (the \textbf{Offline Critical Scenario Detection} phase).
The resulting set of identified failures is filtered to remove redundancy and communicated to the \emph{stakeholders} of the \ac{PDP} system; these comprise professionals involved in both the clinical and organizational dimensions of care delivery, including clinicians and department directors, as well as individuals responsible for training medical staff or overseeing their performance~\cite{fottler1989assessing}.

Leveraging the generated models, our methodology synthesizes action strategies to support physician decision-making in the scenarios under scrutiny.
These strategies can be integrated into a \ac{CDSS}, constituting the \textbf{Online Clinical Decision Support} phase of the methodology, although no \ac{CDSS} is deployed in this work.
In a real-world setting, such a system would interface with the physical world by acquiring patient data, aligning the \ac{DT}'s state with the physical one, and actuating or suggesting clinical actions.

\subsection{Model Definition}
\label{sec:model-definition}
This phase produces an \ac{SHA} representation of the \ac{PDP} triad via a manual \comp{Design} activity and an automated \comp{Learning} activity.
The \comp{Data Model Definition} and \comp{Data Pre-processing} activities must be carried out beforehand to provide well-formed inputs to the learning process.
Finally, \comp{Model Post-processing} refines the generated models to ensure their compatibility with the verification procedures employed in the subsequent phase.

\subsubsection{Data Model Definition}
As we make use of \ac{SHA} (see \sref{sha}), the input required for the learning process includes the set of events $\Sigma$ and the sets of physical quantities to be modeled.
The selection is guided by domain expertise and informed by the operational capabilities and constraints of the available medical equipment, ensuring the learned model captures clinically meaningful and observable behaviors.

Among the selected variables, those that are continuously monitored and require accurate estimation ($X_\mathrm{F}$) are modeled by flow conditions ($\mathrm{F}$).
Other variables ($X_\mathrm{M}$) are primarily informative when they signal specific conditions of the patient or the surrounding environment, such as the violation of critical thresholds that typically trigger alarms on medical equipment, and are thus used for event identification.
The labeling function $\mathcal{L}_{X_\mathrm{M}}$ maps scenario-dependent conditions on variables in $X_\mathrm{M}$ to events in $\Sigma$, thereby formally defining how real-world events are detected from data.
Domain knowledge is essential here for identifying clinically relevant events through possibly complex activation criteria defined over the variables.

\begin{example}
\label{ex:data-model-def}
Consider our running example in \sref{application}. 
Events represent either a significant variation in the patient's vital signs or a change in ventilator settings performed by the intensivist.
Accordingly, these quantities populate the set $X_\mathrm{M}=V\cup P$, where $V=\{\param{CD}, \param{HR}, \param{OS}, \param{RR}, \param{TV}\}$ contains the patient's vitals and $P=\{\param{FIOX}, \param{PEEP}, \param{RERA}, \param{TVOL}\}$ includes the ventilator parameters.
Each patient vital $v_i, i = 1,\dots,5$, is associated with a safe operating range $R_i =[v_{i}^\mathrm{min}, v_{i}^\mathrm{max}]$.
An event is observed when the value of vital $i$ at time $t_k$ (denoted as $v_i^{t_k}$) exceeds its safe range, indicating a deviation from normal physiological conditions, or when it returns within such range, signaling a possible recovery or stabilization.
The ventilator parameters $p_j$, with $j = 1, \dots, 4$ (see \tref{ventilator}), trigger events whenever they are manually increased or decreased by the intensivist.
Finally, the events $\param{on}$ and $\param{off}$ correspond to the activation and deactivation of the mechanical ventilation device, respectively.
When the device is inactive, all ventilator parameters are assumed to be set to zero.
We define the partial labeling function $\mathcal{L}_{X_\mathrm{M}}$ as follows:
\begin{multline*}
\mathcal{L}_{X_\mathrm{M}}(t_k) = 
\begin{cases}
    (\param{V}_i)^{\param{high}} & \text{if } v_i^{t_k} > v_i^{\mathrm{max}} \wedge v_i^{t_{k-1}} \le v_i^\mathrm{max} \\
    (\param{V}_i)^{\param{low}} & \text{if } v_i^{t_k} < v_i^\mathrm{min} \wedge v_i^{t_{k-1}} \ge v_i^\mathrm{min} \\
    (\param{V}_i)^{\param{ok}} & \text{if } v_i^{t_k} \in R_i \wedge v_i^{t_{k-1}} \notin R_i \\
    (\param{P}_j)^{\param{up}} & \text{if } p_j^{t_k} > p_j^{t_{k-1}} \wedge p_j^{t_k}, p_j^{t_{k-1}} \ne 0 \\
    (\param{P}_j)^{\param{down}} & \text{if } p_j^{t_k} < p_j^{t_{k-1}} \wedge p_j^{t_k}, p_j^{t_{k-1}} \ne 0 \\
    \param{on} & \text{if } p_j^{t_k} > 0 \wedge p_j^{t_{k-1}} = 0 \ \forall j \\
    \param{off} & \text{if } p_j^{t_k} = 0 \wedge p_j^{t_{k-1}} > 0 \ \forall j \\
\end{cases} \\
\text{ for } \param{V}_i \in V, \quad \param{P}_j \in P, \quad k = 1, 2, \dots
\end{multline*}
We select $X_\mathrm{F} = \{\param{TV}\}$ as the variable modeled through flow conditions, since even minor variations in its value are critical for assessing the patient's ability to breathe.
\end{example}

\subsubsection{Data Pre-processing}
Raw input data are generally unsuitable for learning algorithms, as meaningful information is not immediately available and must therefore be extracted with proper processing.
Smoothing noisy signals, resampling data to enforce uniform time intervals, or extracting salient features, such as peak values or rates of change, is customary to capture underlying physiological dynamics better.

\begin{example}
In our running example, we apply a smoothing transformation to the $\param{CD}$ (carbon dioxide) signal, which naturally exhibits a sinusoidal breathing-induced pattern.
The raw signal is transformed into a time series that records, at each time step, the most recent peak value observed.
This peak-based representation attenuates high-frequency oscillations while preserving clinically relevant trends, yielding a signal that is more robust to noise and better aligned with the threshold-based event definitions used by the learning activity.
\end{example}

\subsubsection{Design}
Through structured interactions with domain experts (\eg interviews), engineers elicit specifications of the interventions the physicians perform, the triggering events, and the contextual conditions required for each action.
The collected knowledge allows them to formalize components of the \ac{PDP} system by encoding established clinical practices for managing a patient’s evolving health conditions, effectively capturing medical expertise.
Their activity results in models for both the physician ($M_\mathrm{Ph}$) and the ventilation device ($M_\mathrm{D}$).

These elements are specified as a \ac{SHA}, where interventions are modeled as actions on edges and care activities are represented as locations within the physician and device models.
Guards on edges encode the contextual conditions required for executing each action.
When multiple interventions are clinically admissible while the physician is treating the patient, $M_\mathrm{Ph}$ incorporates probabilistic edges with predefined weights, allowing the physician–device interaction model to capture the inherent variability of clinical decision-making.

The resulting \ac{SHA} models are parameterized to represent the degrees of freedom in the intensivist's decision-making process.
Model parameters, defined jointly by engineers and the domain experts, govern key behavioral aspects of the triad and explicitly encode elements of human judgment and uncertainty.
Different parameter instantiations may yield adverse or unsafe patient behaviors, making these parameters central to the Offline Critical Scenario Detection phase.

\begin{example}
\label{ex:design}
\begin{figure}[tb]
    \centering
    \includegraphics[width=.9\linewidth]{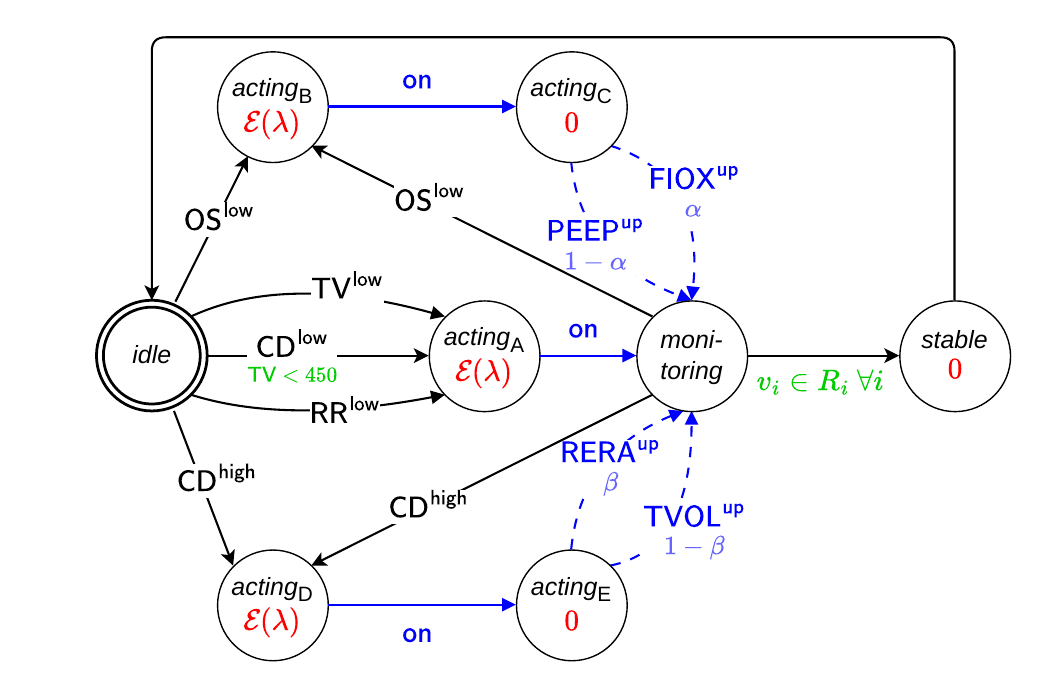}
    \caption{Example of physician-device \ac{SHA}.}
    \label{fig:doctor}
\end{figure}
Figure~\ref{fig:doctor} shows a \ac{SHA} obtained through expert elicitation.
It models the behavior of an intensivist who reacts to clinical events (depicted with black labels on arrows) by activating the mechanical ventilator and adjusting its operational parameters.
Physician-controlled edges are depicted with blue arrows and labels, while stochastic choices are represented by dashed edges annotated with probability weights.
Guards are highlighted in green, and the distributions governing leaving times from locations are shown in red.

According to the model, the physician activates the ventilator (action $\param{on}$) when a decrease is detected in any of the vital signs $\param{TV}$, $\param{CD}$, or $\param{RR}$, provided the ventilator is inactive.
In response to low oxygen saturation ($\param{OS}$) or elevated carbon dioxide ($\param{CD}$), the physician activates the ventilator and adjusts one of its parameters.
For low $\param{OS}$, the physician probabilistically increases either $\param{FIOX}$ or $\param{PEEP}$, with weights $(\alpha, 1-\alpha)$; for high $\param{CD}$, the choice is between increasing $\param{RERA}$ or $\param{TVOL}$, with probabilities $(\beta, 1-\beta)$.
The edge labeled $\param{CD}^\param{low}$ has a guard $\param{TV}<450$, ensuring the physician responds to a $\param{CD}^\param{low}$ alarm only when tidal volume is below 450 mL.
Once all vitals $v_i$ return to their safe ranges $R_i$, the patient is considered stabilized, prompting a transition to the \textit{stable} state, after which the physician returns to the \textit{idle} state.

The \ac{SHA} models realistic response delays for the physician using an exponentially distributed random variable with rate $\lambda$, while other actions are executed instantaneously.
Observation states such as \textit{idle} and \textit{monitoring} lack physician-controlled outgoing transitions; exits from these states occur solely through patient-generated events, so no delay distribution is assigned.
Model parameters are collectively denoted $\varphi = (\alpha, \beta, \lambda)$, where $\alpha \in [0,1]$, $\beta \in [0,1]$, and $\lambda \in [1/15, 1]$%
\footnote{$\alpha$ and $\beta$ are probabilities, while the range for $\lambda$ reflects realistic average ICU response times (in seconds), as informed by domain expertise.}%
.
\end{example}

\subsubsection{Learning}
A data-driven model of patient behavior is essential to captures the evolution of physiological conditions over time and under clinical interventions.
The patient model $M_\mathrm{Pa}$ complements the physician and device models, thereby completing the \ac{PDP} triad.

Patient health conditions in \acp{ICU} are inherently complex and only partially observable, especially in the case of respiratory diseases whose underlying physiological mechanisms cannot be measured directly.
Intensivists therefore rely on indirect indicators, such as vital signs acquired through monitoring equipment, to assess patient status and guide interventions.
This partial observability, combined with inter-patient variability and stochastic physiological responses, motivates the learning of patient dynamics from empirical observations.

The \comp{Learning} activity builds upon the artifacts defined during the data model definition phase, including the set of events, the physical variables, and the labeling function.
It integrates raw physiological signals collected in the \ac{ICU}, which may originate from medical devices, hospital information systems, or manual entries by healthcare professionals.
To infer the patient model, \nameExt{} employs the automata learning algorithm \lsha{}, which derives an \ac{SHA}-based representation from clinical data.
The algorithm infers both the discrete structure of the model (locations and edges) and the continuous dynamics associated with each location.
Specifically, each learned location is equipped with a flow condition governing the evolution of the physical variables in $X_\mathrm{F}$, together with a parametric probability distribution capturing stochastic variability.
Locations thus correspond to distinct physiological states inferred from data, while edges represent clinically relevant events that drive changes between such states.

\begin{example}
Figure~\ref{fig:patient} illustrates a portion of the patient \ac{SHA} model learned using \lsha{}.
\begin{figure}[tb]
    \centering
    \includegraphics[width=.7\linewidth]{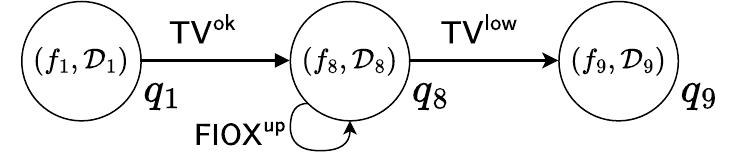}
    \caption{Detail of learned patient \ac{SHA} having $11$ locations.}
    \label{fig:patient}
\end{figure}
The flow condition $f_i$ and stochastic parameter distribution $\mathcal{D}_i$ jointly govern the evolution of variable ${\param{TV} \in X_\mathrm{F}}$ while the system resides in location $q_i$.
From location $q_1$, the occurrence of the $\param{TV^{ok}}$ event, indicating that tidal volume has returned to safe levels, moves the patient to location $q_8$.
From there, event $\param{TV^{low}}$ transitions the patient to location $q_9$.
The $\param{FIOX^{up}}$ self-loop represents that the physician's action of increasing $\param{FIOX}$ has no observable effect on the patient's state while in $q_8$.
\end{example}

\subsubsection{Post-processing}
The final activity of model definition combines the individual models of the \ac{PDP} components (namely physician $M_\mathrm{Ph}$, device $M_\mathrm{D}$, and patient $M_\mathrm{Pa}$) into a single, unified model representing the triad.
This composite model is further refined to ensure compatibility with the verification procedures of the subsequent phase.
The refinement produces the final model $M$ as output.

\begin{example}
The patient model $M_\mathrm{Pa}$ learned from data is inherently incomplete, as not all locations necessarily contain outgoing edges for every event in $\Sigma$.
While manually designed models can explicitly handle all possible events in every location, learned models typically reflect only the behaviors observed during training.
Some events may be left untreated in some locations if they never occurred during data collection.

Under such circumstances, typical post-processing includes ensuring the model is well-defined for all events, by introducing a \emph{sink} location representing unmodeled or unsupported behaviors for which data is insufficient.
For each location in $M_\mathrm{Pa}$ lacking an outgoing edge for some event $e \in \Sigma$, an edge labeled with $e$ is added and linked to the sink location.
\end{example}

\subsection{Offline Critical Scenario Detection}
\label{sec:sub:scenario-detection}

This phase evaluates whether the \ac{SHA} digital model $M$ satisfies the dependability requirements required by stakeholders.
The \comp{Model space analysis} activity leverages an initial specification of formal requirements to systematically explore variations of model parameter and structure. Candidate generation combined with formal verification enable the identification of failure-inducing configurations.
Clinically unrealistic scenarios are filtered with the \comp{Validation} activity, and through the \comp{Diversity analysis} activity representative and meaningful critical scenarios are selected.

\subsubsection{Dependability Requirements Specification}
To identify and formalize the system properties guaranteeing patient safety and reliable treatment, engineers directly collaborate with stakeholders.
The latter may include clinicians, who define relevant medical conditions; biomedical engineers, who specify equipment constraints; safety analysts, who ensure compliance with risk mitigation standards; and department directors, who focus on operational efficiency.
This elicitation process allows domain experts to contribute expectations and operational constraints, which are then translated into verifiable properties.
The specification language must support properties with a probabilistic interpretation over \ac{SHA} models, enabling reasoning about uncertainty and temporal behavior in \ac{PDP} triads.

The activity yields a set $\Psi$ of dependability properties, each of which can be evaluated on a given \ac{SHA} instance as satisfied or violated with an associated probability.
Such probabilistic semantics enables the identification of scenarios in which dependability degrades under uncertainty.
The activity also defines a reachability goal $G$, specified as a set of safe automaton states and a cost metric to minimize.
Reachability of these states constitutes a dependability objective that will be exploited in the online clinical decision support phase.

\begin{example}
\label{ex:reqs}
Consider again our running example presented in \sref{application}. 
We define three dependability properties evaluated over a fixed time horizon:
\begin{enumerate}
    \item \textit{Reachability of patient stability}: reaching a location representing the stabilization of patient vitals;
    \item \textit{Duration in non-breathing state}: $\param{TV}$ (tidal volume) remains below a defined threshold for a specified duration;
    \item \textit{Persistence of critical health condition}: remaining in a state with multiple vitals outside their safe ranges for a sustained period.
\end{enumerate}
These properties are expressed in \ac{MITL}~\cite{david2011statistical} formulae.
The probability that the first formula holds is expected to fall below a stakeholder-defined threshold, whereas the probabilities for the second and third are evaluated to exceed their respective thresholds.
Traces reaching the sink location are not considered violations, as this represents unmodeled behaviors and an unknown system state.
Alternative interpretations (e.g., treating these behaviors as failures) are possible and can be implemented by adjusting the \ac{MITL} properties.
We define goal $G$ as the singleton set containing the location of the patient model representing a stable condition and the time spent in that location as the cost metric, which we aim to maximize.
\end{example}

\subsubsection{Model Space Analysis}
Automated techniques are adopted to systematically explore the \ac{PDP} model space and identify diverse critical scenarios.
Model space analysis follows an iterative process alternating between \comp{Space Exploration} and \comp{Verification}.
As introduced above, the \ac{PDP} model $M$ is parameterized by vector $\varphi$, representing the degrees of freedom in the behavior of both the physician and device models.
The exploration step is realized either by selecting an assignment to parameters $\bar\varphi$ or by changing structural features of $M$, yielding a modified model $M'$,
while the verification step evaluates the probability that $M'[\bar\varphi]$ satisfies the defined dependability requirements $\Psi$.
These probabilities are fed back into the exploration step in a feedback loop.
Specifically, the probability that models satisfy $\Psi$ is computed through \acf{SMC}~\cite{legay2010statistical}.
If the computed probability of any undesirable behavior exceeds a stakeholder-defined threshold, the model is considered a \emph{failure}.
During model space exploration, failure-inducing \acp{PDP} models (or more precisely, their parameterizations $\varphi$) are collected in a set $\Phi$ to identify critical behaviors of the physician-device system that may compromise patient safety.
These failures highlight specific combinations of clinical decisions and system responses that merit further investigation, either as indicators of potential safety risks or opportunities for system improvement.

\nameExt{} supports two complementary exploration strategies: mutational \emph{fuzzing} and \emph{search-based} optimization (see \sref{background}).
Both strategies modify model parameters or structural elements of the \ac{SHA} model and are guided by the likelihood of a model violating requirements as determined during \comp{Verification}.
\emph{Mutational fuzzing} applies domain-specific mutation operators to introduce behavioral variations into the \ac{SHA}, iteratively generating new \emph{mutants}.
These operators maintain model validity while producing meaningful variability.
An archive of physician-device \ac{SHA} mutants is maintained and enriched with new mutants added according to application-specific criteria, serving as a pool of promising candidates for further mutation.
By prioritizing high-risk mutants, the algorithm directs exploration toward underexplored regions of the model space, increasing the likelihood of uncovering critical behaviors.
The \emph{search-based} strategy treats model space exploration as a multi-objective optimization problem, with each objective minimizing the probability of satisfying a specific dependability property.
Unlike fuzzing, the definition of the candidates to evaluate is not generally driven by application-dependent criteria, but rather depends on the employed algorithm, which introduces variability in an application-agnostic manner (still by acting on model parameters and on the structural elements of the automata).
These strategies produce a collection of physician-device variants, each representing a distinct instantiation of model $M$, which are then analyzed via verification and potentially labeled as failures.

\begin{example}
\label{ex:space-analysis}
In our running example, we define two mutation operators for fuzzing.
The first perturbs a randomly selected parameter in $\varphi$ by applying a bounded scaling factor, thereby modifying the intensivist's reaction times or probabilistic preferences when selecting interventions in the \ac{SHA} shown in \fref{doctor}.
The second removes a randomly chosen transition, modeling situations in which the physician does not respond to a specific clinical event.
We set the criterion for adding a mutant to the archive as exceeding the current maximum observed likelihood of violation for at least one requirement.
An example of an alternative physician–device model removes the topmost transition (labeled by $\param{on}$) that activates the ventilator in response to a low oxygen saturation event, while two others increase the physician's reaction time (\eg{} $\lambda = 1/15$ and $\lambda = 2/25$), representing delayed responses due to workload or organizational factors.
\ac{SMC} results indicate a violation of at least one requirement (see Example~\ref{ex:reqs}), leading the corresponding models to be classified as failure-inducing scenarios.
We employ the Non-dominated Sorting Genetic Algorithm II (NSGA-II)~\cite{deb2002fast} as the multi-objective searching algorithm.
Unlike fuzzing, candidate generation is not driven by application-dependent criteria, but is performed in a black-box manner by the algorithm, which perturbs model parameters and structure.
\end{example}

\subsubsection{Validation}
\ac{PDP} mutants in $\Phi$ that yield a failure but represent unrealistic or clinically irrelevant scenarios are filtered out, obtaining a reduced subset of parameters $\Phi' \subseteq \Phi$.
While some of these cases may technically pose risks to patient safety, they are considered either uninteresting for further analysis or unlikely to occur in real-world settings, based on stakeholder judgment.
For example, a case where a physician takes an excessively long time (\eg{} minutes) to respond to an emergency may be deemed unrealistic, given the high availability of personnel in typical \ac{ICU} environments.
Other scenarios may involve implausible clinical actions that no trained physician would reasonably perform.
Because such scenarios cannot always be excluded a priori, they must be explicitly reviewed by clinical experts to assess their validity.

\begin{example}
\label{ex:validation}
After a guided review of the \ac{SHA} components in \fref{doctor}, the expert identifies edges essential for any clinically reasonable procedure.
Based on this review, a set of structural and parametric constraints is derived to serve as exclusion criteria.
These constraints enable automated filtering of scenarios unlikely to occur in real \ac{ICU} settings.
For example, among the three \ac{SHA} models in \exref{space-analysis}, the first is excluded as it lacks an appropriate physician response to the $\param{OS^{low}}$ event, a behavior deemed extremely unlikely by the domain expert.
\end{example}

\subsubsection{Diversity Analysis}
The set of realistic failure-inducing \ac{PDP} models in $\Phi'$ is subjected to a diversity analysis to identify distinct underlying root causes.
Indeed, in some cases, failure-inducing models that differ syntactically may nevertheless exhibit semantic similarity, and hence fail to correspond to distinct scenarios.
To reduce redundant information and obtain a reduced set of representative critical scenarios, unsupervised clustering~\cite{sokal1958statistical} is applied to group similar failure-inducing \ac{PDP} models based on their behavioral characteristics and, from each cluster, a representative model is extracted.
All the representatives collectively form the final set of diverse failure scenarios presented to stakeholders.
This process requires a definition of \emph{similarity} between models and a \emph{selection criterion} for choosing a representative from each cluster.
Both definitions are collaboratively established with domain experts to balance clinical and analytical relevance.
Failure-inducing \ac{PDP} models can be clustered by analyzing the parameter space defined by their associated vectors $\varphi$.

\begin{example}
\label{ex:clustering}
We apply unsupervised hierarchical clustering using the Unweighted Pair-Group Method with Arithmetic mean (UPGMA)~\cite{sokal1958statistical}, which iteratively merges clusters based on the average pairwise distance between all elements across candidate clusters, thereby capturing similarity relationships in the parameter space.
Rather than producing a single partition, UPGMA yields a dendrogram that represents the hierarchical organization of the \ac{PDP} models.
A concrete clustering is then obtained by cutting the dendrogram at a selected distance threshold, which allows the number and granularity of clusters to be tuned to the desired level of abstraction.
In our evaluation, we consider multiple candidate clusterings with different numbers of clusters and automatically select the one that maximizes the silhouette score~\cite{rousseeuw1987silhouettes}, a widely adopted metric for assessing cluster cohesion and separation.

The parameters $\varphi$ of the two remaining example \ac{SHA} in \exref{validation} are almost identical, differing only in the physician's response time.
Accordingly, the clustering procedure considers them to be close in the parameter space and assigns them to the same cluster.
This cluster can be interpreted as the set of critical circumstances due to delayed action by the physician (\ie slow response time).
The centroid of the cluster is then selected as its representative.
\end{example}

\subsection{Online Clinical Decision Support}
This phase completes the implementation of the \ac{DT} of the \ac{PDP} triad by establishing a feedback loop between its physical and the digital copies.
Specifically, the \ac{DT} requires \emph{alignment} between the physical state of the system and its digital representation through a continuous, bidirectional flow of data and corresponding state updates.

A selected model, representing the current status of the physical \ac{PDP} triad in the \ac{DT} loop, is translated into a \ac{PTGA}, and a strategy is computed, ensuring reachability of a prescribed goal.
The strategy is then used in \comp{Decision Selection}, while state alignment (highlighted in blue in \fref{overview}) is maintained via \comp{Digital State Identification} and \comp{Physical State Actuation}.
The models for which the control strategy is calculated can be either the representatives produced by the diversity analysis, or generalist models that do not capture specific criticalities of the physical triad but instead represent possible nominal operating scenarios. The latter can be developed starting from the model obtained through the activity described in Section \ref{ex:data-model-def} or from its generalizations.

\subsubsection{Controller Synthesis}
Since patient safety is a primary concern, a controller is synthesized to assist the physician through recommended actions that aim at preserving patient stability.
The methodology leverages game-theory techniques for which an entirely stochastic \ac{SHA} model is interpreted as a 1\textonehalf-player game, modeling the physician (\ie the \emph{player}) as a goal-oriented strategy-based player and the patient (\ie the \emph{adversary}) as a stochastic adversary~\cite{david2015uppaal}.
Specifically, the \ac{SHA} model is converted into a \ac{PTGA} where the physician's actions, instead of being probabilistic as in the \ac{SHA}, are governed by a strategy while the patient's reactions are stochastic.
Hence, the \ac{SHA}-to-\ac{PTGA} conversion leaves the patient model unvaried and marks all physician's edges as controllable.

Based on a formal \ac{PTGA} representation of the system, derived from a selected model $M[\bar\varphi]$, the \comp{Controller Synthesis} activity computes a strategy $\controller_{M[\bar\varphi]}$ that prescribes the optimal action at each possible system state to satisfy the reachability goal $G$ specified by stakeholders.
The strategy serves as a physician decision-support engine in the \ac{DT} alignment loop.

\begin{example}
Consider a modified version of the physician-device \ac{SHA} of \fref{doctor} with localized changes in each \emph{acting} and \emph{monitoring} location, represented in \fref{doctor-strat}.
\begin{figure}[tb]
    \centering
    \includegraphics[width=.9\linewidth]{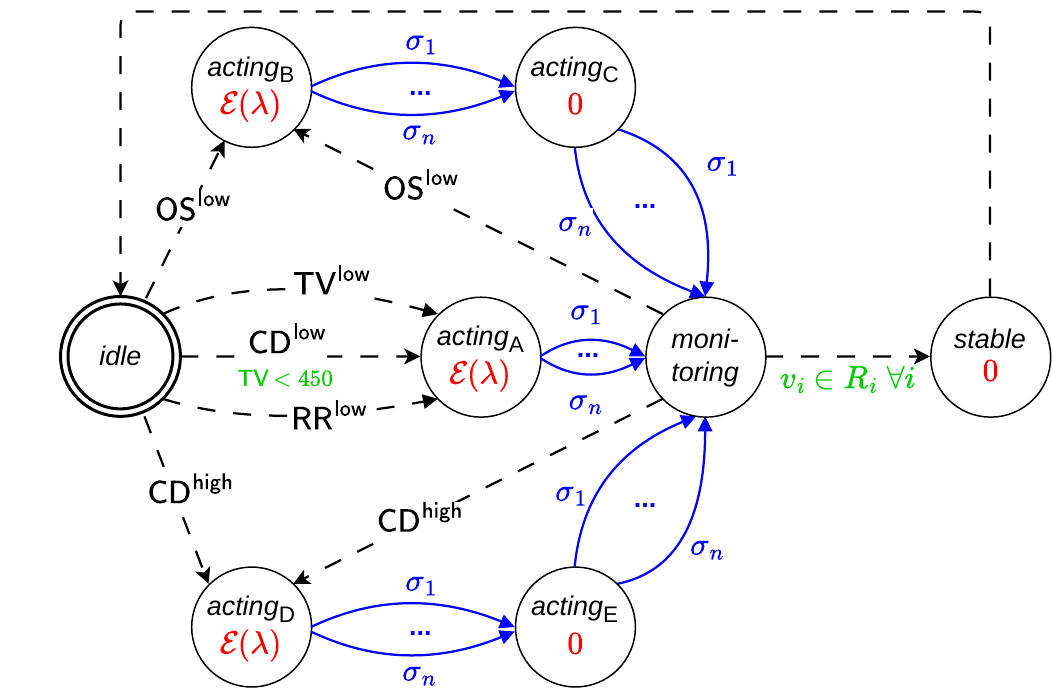}
    \caption{Physician-device \ac{PTGA} used in the running example. Solid blue edges are the physician's controllable actions $\sigma_1, \dots,\sigma_n \in \Sigma_\mathrm{o, c}$. Dashed edges are uncontrollable actions.}
    \label{fig:doctor-strat}
\end{figure}
This model is a generalization of the physician model shown in \fref{doctor} that does not encode any specific medical procedures, yet it represents a generalized physician with full freedom to adjust the ventilator parameters.
Specifically, it includes outgoing edges for all the physician-controlled actions $\sigma_1$, \dots, $\sigma_n$ in $\Sigma_\mathrm{o, c}$, where $\Sigma_\mathrm{o, c} := \{(\param{P}_j)^{\param{up}}, (\param{P}_j)^{\param{down}}\}_{\param{P}_j \in P} \cup \{\param{on}, \param{off}\}$.
By definition, the \ac{PTGA} configuration tuple $(l, \nu)$ includes:
\begin{enumerate*}[\textit{\roman*})]
    \item the current locations of both automata;
    \item a collection of boolean variables $r_i$, with ${\param{V}_i \in V}$, indicating whether each patient's vital currently falls within ($\param{T}$) or outside ($\param{F}$) its safe range $R_i$;
    \item a boolean variable $r_\param{on}$ indicating whether mechanical ventilation is currently active ($\param{T}$) or inactive ($\param{F}$).
\end{enumerate*}
The learned controller $\controller_{M[\bar\varphi]}$ maximizes permanence time in {the location defining goal $G$ (see Example~\ref{ex:reqs})}.
One illustrative tuple-action mapping of the learned strategy is:
\[\controller_{M[\bar\varphi]}(acting_\mathrm{A}, q_9; \param{T}, \param{T}, \param{T}, \param{T}, \param{F}, \param{F}) = \param{on}.\]
The mapping states that if the physician-device \ac{SHA} is currently in location $acting_\mathrm{A}$, the patient \ac{SHA} is in location $q_9$, the first four patient vitals are stable while the fifth one ($\param{TV}$) is not, and mechanical ventilation is inactive, the action suggested by the strategy is $\param{on}$, \ie activate the ventilator.
\end{example}

\subsubsection{Digital State Identification}
Within the DT paradigm, a digital representation of the current state of the physical PDP triad, captured in the SHA model $M[\bar\varphi]$ via locations and configuration variables, is essential for enabling control feedback through recommendations and formal analysis.
State identification can be achieved by:
\begin{enumerate*}[\textit{\roman*})]
    \item acquiring the patient's vital signals in real time, including $X_\mathrm{M}$;
    \item applying labeling function $\mathcal{L}_{X_\mathrm{M}}$, which produces a trace $\tau$ representing the events recorded in the physical environment; and
    \item interpreting $\tau$ on the \ac{PDP} model.
\end{enumerate*}
The location reached in the \ac{SHA} by sequentially firing the events from $\tau$, and the sampled vitals, represent the digital counterpart of the current state of the physical system.
The initial location of the model at the onset of the feedback loop is the one marked as \textit{initial} in the \ac{SHA}.

\begin{example}
Consider the representation of the \ac{PDP} triad shown in \fref{patient} (partially) for the patient and \fref{doctor-strat} for the physician.
Let us assume that, at a certain time instant, the patient's current location is $q_8$ and the physician's current location is \textit{idle}, and that readings from the patient's tidal volume indicate that the metric has dropped below the critical threshold $\param{TV}^\mathrm{max}$.
The labeling function indicates that event $\mathcal{L}_{X_\mathrm{M}}(t_k) = \param{TV}^\param{low}$ has taken place.
The patient and physician \ac{SHA} transition to locations $q_9$ and $\textit{acting}_\mathrm{A}$, respectively.
\end{example}

\subsubsection{Decision Selection}
The actions implementing the control feedback loop are identified by means of a \textit{decision engine}, which is based on the controller $\controller_{M[\bar\varphi]}$ synthesized during the previous phase.
Given the current system state $(l, \nu)$, the engine selects an \ac{SHA} event $\sigma = \controller_{M[\bar\varphi]}(l, \nu) \in \Sigma_\mathrm{o, c}$.
This event is the formal representation of a physician's action and corresponds to the behavior that results in {minimal cost when reaching goal $G$}.
The selected action is sent to the physical state actuator either as a recommendation presented to the physician through a digital interface or as a direct control signal to an automated device.
Concurrently, event $\sigma$ is fired in the \ac{SHA} network, potentially triggering a location update in the patient and/or physician automata.

\begin{example}
The synthesized strategy prescribes the activation of the ventilator, corresponding to the firing of event $\sigma = \param{on}$.
As a consequence, the physician \ac{SHA} transitions from location $\textit{acting}_\mathrm{A}$ to \textit{monitoring}, and a corresponding update occurs in the patient \ac{SHA} if the current state $q_9$ admits an outgoing transition labeled with event $\param{on}$.
\end{example}

\subsubsection{Physical State Actuation}
The actuation of the decision engine's output in the physical world can take two forms.
It can be fully automated, with the actions sent directly to the device, which updates its settings accordingly and in real time.
Alternatively, they can be provided as a recommendation to the physician, leaving the final decision to the physician’s judgment and expertise.
In either case, the decision affects the patient's physiological state and metrics, a process corresponding to the second half of the \ac{DT} alignment loop.

Physical state actuation closes the \ac{DT} alignment loop, as it enables decisions produced in the digital domain to influence the physical system.
It takes place entirely in the real world and depends on deployment-specific considerations, such as device interoperability, clinical workflows, and regulatory constraints.
For this reason, the concrete implementation of actuation, whether fully automated or mediated by the physician, is beyond the scope of this work, which focuses on the methodological foundations for constructing and analyzing \acp{DT}.
Accordingly, in \fref{overview} this activity is depicted as an external block positioned outside the third phase of the methodology.

\section{Empirical Evaluation}
\label{sec:exp}

This section presents the experimental campaign\footnote{Replication package found at \url{https://doi.org/10.5281/zenodo.18430307}.} designed to assess both the accuracy and the scalability of the proposed methodology.
The evaluation is structured around the following research questions:
\begin{enumerate}[label=\textbf{RQ\arabic*:}, leftmargin=28pt]
    \item What is the accuracy of the patient model resulting from the learning phase of \nameExt{}?
    \item What is the effectiveness of \nameExt{} in detecting realistic and diverse failure scenarios?
    \item What is the cost of the failure detection phase of \nameExt{}?
    \item What is the effectiveness of the strategies synthesized by \nameExt{} in stabilizing patient metrics?
\end{enumerate}

\subsection{Design of the evaluation}

\subsubsection{Evaluation subjects}
Synthetic patient data are generated using Kitware \textsc{Pulse}~\cite{bray2019pulse}, an open-source high-fidelity physiological simulator capable of reproducing diverse clinical conditions and adverse events.
The simulations are executed through BREATHE~\cite{colombo2025breathe}, a framework built on \textsc{Pulse} that provides a clinician-oriented interface for real-time monitoring of vital signs and interactive control of ventilator settings.

Data collection to train the predictive models was carried out with the support of a physician with four years of experience in intensive care medicine.
The clinician conducted $8$ simulated clinical scenarios using BREATHE, for a cumulative simulation duration of $45$ minutes.
The executed scenarios covered $10$ distinct health complications, together with the corresponding \ac{ICU} interventions and resulting patient responses.
During the simulations, the intensivist was provided with real-time physiological data and adjusted ventilation parameters according to their clinical judgment and established medical guidelines.

The physician also contributed to the definition of our main study subject, that is, the respiratory intensive care example introduced in Section~\ref{sec:application} and Section~\ref{sec:methodology}, by identifying relevant physiological signals, describing medical procedures with \ac{SHA}, and specifying validation criteria to discriminate clinically realistic scenarios from unrealistic ones.

\subsubsection{Methods under comparison}
We assess the cost-effectiveness of \nameExt{} by comparing it against multiple baseline methods, selected according to the specific RQ.

To evaluate the accuracy of the learned patient model (RQ1), we compare it with three mainstream regression baselines.
All models share the same input-output specification: the input encodes a simulation trace as a sequence of events (set $\Sigma$ as per \ref{sec:model-definition}), represented as categorical variables, together with the corresponding $\param{TV}$ values associated with each event.
The output is a single scalar representing the predicted $\param{TV}$ after the last event in the trace.

The considered baseline models are: \textit{dual-input} \ac{NN}~\cite{goodfellow2016deep}, \textit{XGBoost}~\cite{chen2016xgboost}, and \textit{Elastic Net}~\cite{zou2005regularization}.
The \textit{dual-input} \ac{NN} uses one branch to encode the categorical event sequence through an embedding layer followed by an LSTM layer~\cite{hochreiter1997long}, while the second branch processes the $\param{TV}$ values using a convolutional layer followed by pooling.
The outputs of the two branches are concatenated and passed through a fully connected layer with dropout to mitigate overfitting.
The \textit{XGBoost} is a gradient-boosted decision tree model operating on a combined input vector that includes both the encoded event sequence and the corresponding $\param{TV}$ values.
The \textit{Elastic Net} is a linear regression model with combined $L^1$ and $L^2$ regularization (same input representation as XGBoost).

We evaluate the effectiveness (RQ2) as well as the cost (RQ3) of \nameExt{} in detecting realistic and diverse failure scenarios by comparing alternative strategies for exploring the \ac{PDP} model space.
Specifically, we compare the proposed approaches---\emph{mutational fuzzing} and a \emph{search-based} strategy based on NSGA-II---against a \emph{random search} baseline, which uniformly samples random mutations over both model parameters and structural elements.

Concerning the effectiveness of the strategies to stabilize the patients (RQ4), we compare \nameExt{} against a ground-truth decision-making process of a real human physician.

\subsubsection{Statistical tests}
We reduce the risk of obtaining results by chance evaluating $20$ distinct scenarios and executing each exploration strategy $20$ times under the same budget constraints (\ie{} producing a total of $500$ patient model mutants).
Preliminary tests indicate that this budget is adequate to reach a performance plateau during model space exploration.
Following the guideline by Arcuri and Briand~\cite{ArcuriICSE2011}, we employ the non-parametric Mann–Whitney U test~\cite{MannWhitney} to assess the statistical significance of our results.
Additionally, we calculate Vargha and Delaney's \vd{} score~\cite{VDA} to measure the effect size between two samples, offering an indication of stochastic dominance.
We categorize the effect size \vd{} ($= 1- \hat{A}_{21}$) using the following standard classes: small (S), medium (M), or large (L) when its value is at least $0.56$, $0.64$, or $0.71$, respectively; none (N) otherwise.

\subsubsection{Evaluation testbed}
All experiments were executed on an Ubuntu 24.10 system with standard hardware: an 8-core Intel Core i5 processor at 4.2GHz and 16 GB of RAM.

\subsection{Evaluation results}

\subsubsection{Model accuracy (RQ1)}
We assess the accuracy of the model learned with \lsha{} by observing the variable governed by flow conditions, specifically $\param{TV} \in X_\mathrm{F}$, in the patient \ac{SHA}.
The test set consists of 20 scenarios encompassing health complications of different types and severities, along with a sequence of events describing the onset of a particular health complication and randomly chosen physician actions.
Each scenario is simulated in BREATHE to obtain the actual final $\param{TV}$ values as ground truth that we used to compare the predicted $\param{TV}$ value obtained from the learned patient \ac{SHA}, coupled with the complete physician-device \ac{SHA} from \fref{doctor}, and the trained baseline regressors.
The \ac{SHA} modeling \ac{PDP} triads are translated into \uppaal{} models~\cite{behrmann2004tutorial}, from which we extract the predicted $\param{TV}$ at the end of each trace.

We evaluate our \ac{PDP} model against the baselines on two prediction tasks: \emph{regression} and \emph{classification}.
For the \emph{regression} task, we calculate each model's test-set \ac{MAPE} relative to the ground truth.

\begin{figure}[tb]
    \centering
    \includegraphics[width=\linewidth]{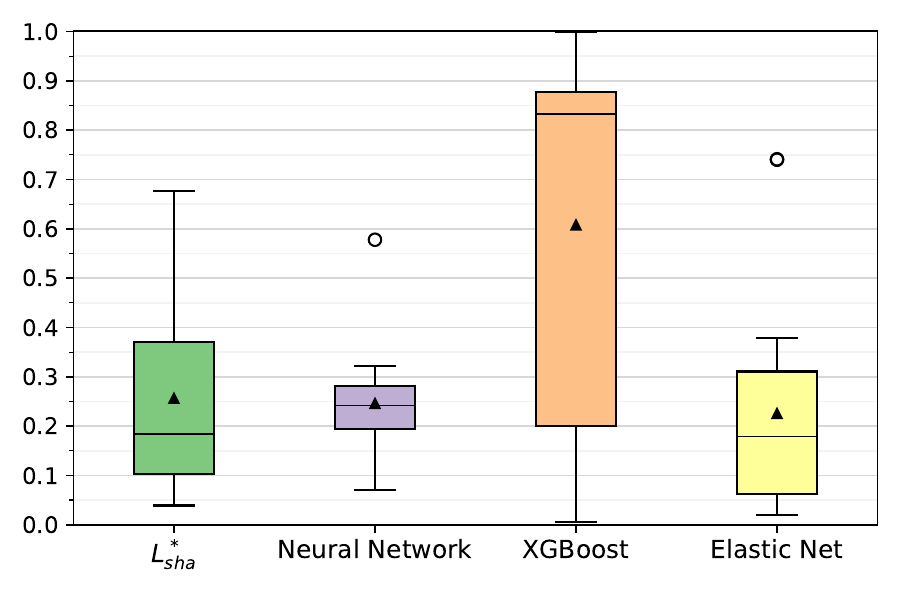}
    \caption{RQ1 \ac{MAPE} values for each regression model.}
    \label{fig:regression-boxplot}
\end{figure}

\vspace{.3em}\noindent\textbf{Results}.
Figure~\ref{fig:regression-boxplot} shows the distribution of \ac{MAPE} for each model, while \tref{regression-stat} provides the Mann–Whitney $p$-values and Vargha–Delaney effect sizes (\vd{}).
For \emph{classification}, we discretize the regression outputs into three categories $(low, ok, high)$ based on their position relative to $\param{TV}$'s safe operating range, as defined in \exref{data-model-def}.
Classification accuracy is then computed as the proportion of correctly classified samples.
\tref{rq1-metrics} collects relevant metrics for both prediction tasks, namely median \ac{MAPE}, mean \ac{MAPE}, and classification accuracy.

\begin{table}[tb]
\caption{Results of statistical tests for regression (RQ1).}
\label{tab:regression-stat}
    \centering
    \begin{tabular}{lcc}
        \toprule
        & \mw{} & \vd{} \\
        \midrule
        \lsha{} vs Neural Network & 3.97e-01 & S \\
        \lsha{} vs XGBoost & 9.35e-03 & L \\
        \lsha{} vs Elastic Net & 4.83e-01 & S \\
        Neural Network vs XGBoost & 2.27e-02 & L \\
        Neural Network vs Elastic Net & 2.09e-01 & S \\
        XGBoost vs Elastic Net & 5.54e-03 & L \\
        \bottomrule
    \end{tabular}
\end{table}

\begin{table}[tb]
\caption{Scores for regression and classification (RQ1).}
\label{tab:rq1-metrics}
    \centering
    \begin{tabular}{l|c|c|c}
        \toprule
        & \multicolumn{2}{|c|}{regression (\ac{MAPE})} & classification \\
        \midrule
        model & median & mean & accuracy \\
        \midrule
        \lsha{}        & 0.184 & 0.258 & 0.90 \\
        Neural Network & 0.241 & 0.247 & 0.40 \\
        XGBoost        & 0.833 & 0.609 & 0.30 \\
        Elastic Net    & 0.179 & 0.227 & 0.70 \\
        \bottomrule
    \end{tabular}
\end{table}

The \lsha{} model consistently outperforms XGBoost in both regression and classification tasks.
Its average \ac{MAPE} is similar to that of the \ac{NN} and Elastic Net models, though statistical testing does not allow us to reject the null hypothesis that these three models share the same \ac{MAPE} distribution.
We can observe that \lsha{} achieves substantially higher classification accuracy, more than twice that of the \ac{NN}, suggesting a deeper understanding of the underlying phenomenon.

\begin{rqbox}[RQ1 summary.]
The \ac{SHA} learned by \lsha{} outperforms all three baseline models: dual-input \acf{NN}, XGBoost, and Elastic Net.
This advantage is especially clear in classification, where \lsha{} reaches accuracy levels up to $3\times$ higher.
For regression, \lsha{} performs on par with the \ac{NN} and Elastic Net, while reducing average error by $3$–$4\times$ relative to XGBoost.
\end{rqbox}

\subsubsection{Effectiveness of failure detection (RQ2)}
We measure effectiveness by counting the occurrences of \emph{different} failure-inducing scenarios discovered during \ac{PDP} space exploration, along with evaluating their realism.
Using the complete physician–device model shown in \fref{doctor}, we apply the mutation operators outlined in Section~\ref{sec:methodology}: parameter perturbation with a mutation factor of $k = 1.5$ (while respecting parameter bounds) and edge removal.
All methods are executed under the same computational budget of $500$ generated mutants (corresponding to $50$ generations of $10$ individuals each for the search-based strategy).
For each technique and requirement, we filter out failure cases that are not clinically realistic according to the criteria defined by the domain expert, as described in Section~\ref{sec:methodology} (see \exref{validation}).
We then compute the rate of realistic failure cases over the total budget.

To assess the diversity of the identified failures, we apply the clustering procedure (see \exref{clustering}) to the set of realistic failure cases.
We evaluate clusterings composed of up to $20$ clusters and select the one that maximizes the silhouette score.
The resulting number of clusters is used as a surrogate measure of failure diversity, capturing the extent to which a technique explores distinct regions of the model space and uncovers qualitatively different classes of clinically relevant failures.
For both failure detection rate and number of clusters, higher values indicate better performance, as they reflect a greater ability to uncover a wide variety of critical scenarios.

\begin{figure}[tb]
\centering
\includegraphics[width=\linewidth]{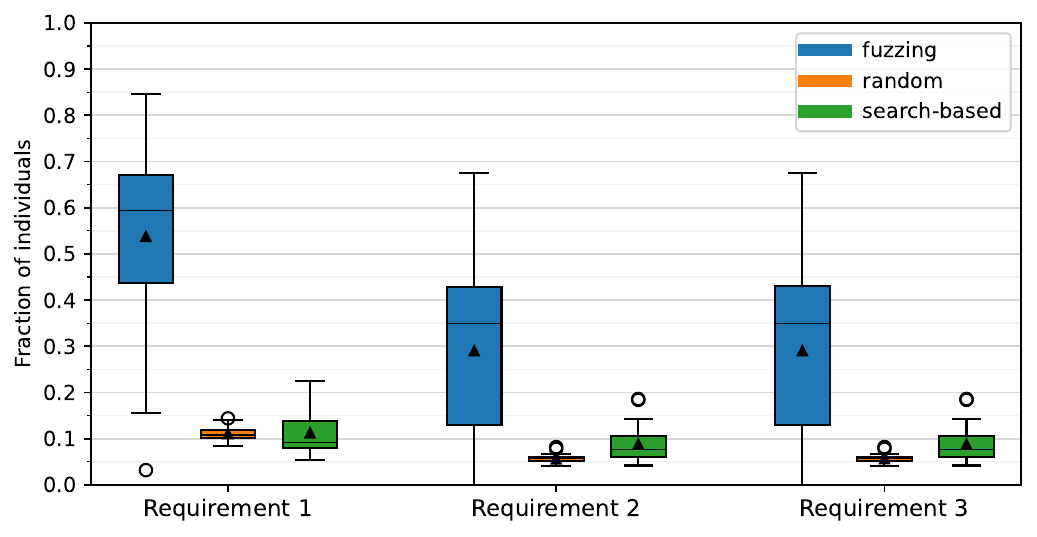}
\caption{Detection rate of realistic failures for each requirement and exploration technique (RQ2).}
\label{fig:testing}
\end{figure}

\vspace{.3em}\noindent\textbf{Results}.
Figure~\ref{fig:testing} shows the distributions of the number of realistic failures identified by each technique over $20$ repetitions.
Similarly, \fref{clustering} reports the distributions of cluster cardinalities across repetitions.
\tref{testing} presents statistical significance and effect size for all pairwise comparisons, considering both the number of realistic failures and the number of clusters.

\begin{figure}[tb]
\centering
\includegraphics[width=\linewidth]{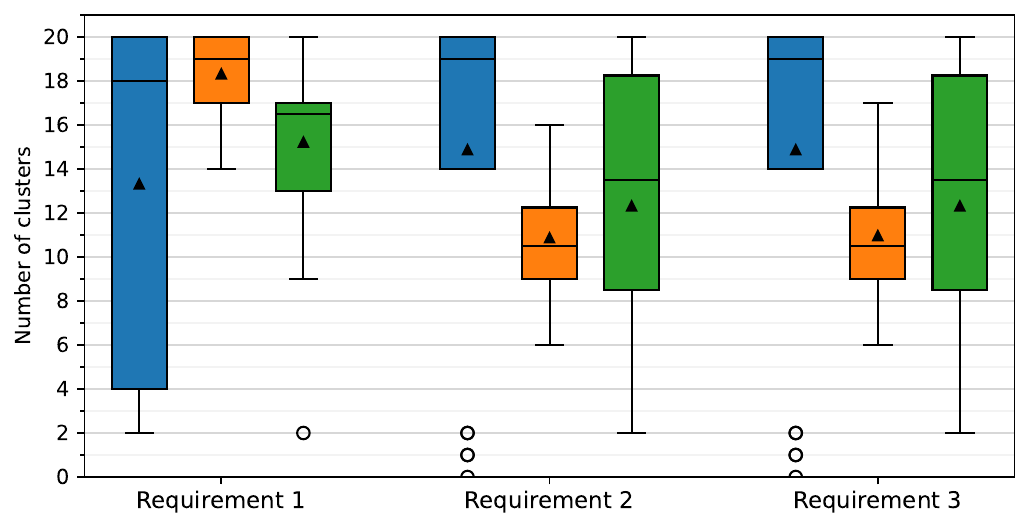}
\caption{Number of clusters for each requirement and exploration technique (RQ2). Color code is the same as \fref{testing}.}
\label{fig:clustering}
\end{figure}

\setlength{\tabcolsep}{3pt}
\begin{table}[tb]
\centering
\caption{Statistical tests for model space exploration and clustering (RQ2).}
\label{tab:testing}
\begin{tabular}{lcc|cc|cc}
\toprule
Realistic failures & \multicolumn{2}{c|}{Requirement 1} & \multicolumn{2}{|c|}{Requirement 2} & \multicolumn{2}{|c}{Requirement 3} \\
\midrule
& \mw{} & \vd{} & \mw{} & \vd{} & \mw{} & \vd{} \\
\midrule
fuzzing vs random & 1.19e-06 & L & 1.22e-03 & L & 1.22e-03 & L \\
fuzzing vs search-based & 2.35e-06 & L & 3.04e-03 & L & 3.04e-03 & L \\
random vs search-based & 1.98e-01 & S & 5.87e-03 & L & 6.86e-03 & L \\
\bottomrule
\toprule
Clustering & \multicolumn{2}{c|}{Requirement 1} & \multicolumn{2}{|c|}{Requirement 2} & \multicolumn{2}{|c}{Requirement 3} \\
\midrule
& \mw{} & \vd{} & \mw{} & \vd{} & \mw{} & \vd{} \\
\midrule
fuzzing vs random & 3.85e-01 & S & 6.71e-03 & L & 6.71e-03 & L \\
fuzzing vs search-based & 6.49e-01 & N & 8.29e-02 & M & 8.29e-02 & M \\
random vs search-based & 2.87e-03 & L & 1.54e-01 & S & 1.62e-01 & S \\
\bottomrule
\end{tabular}
\end{table}
\setlength{\tabcolsep}{6pt}

Mutational fuzzing identifies proportions of realistic failure scenarios ranging between $15$ and $65\%$ across different requirements.
In particular, fuzzing detects failures in approximately $60\%$ of cases (median value) for Requirement $1$ and $35\%$ for Requirements $2$ and $3$.
It also achieves the largest number of distinct clusters, with median values ($18$–$19$) close to the maximum number of clusters evaluated across all requirements.
Statistical tests confirm this superiority in all cases except one, \ie{} the comparison against random search for the number of clusters derived from Requirement $1$.

The failure detection rate of the search-based approach is much lower than fuzzing, around $10\%$ for all requirements.
This trend also holds for the number of clusters, sometimes comparable but lower on average.
In this sense, performance of the search-based approach is comparable to random search.

Overall, the results suggest that fuzzing is the most effective technique for uncovering a larger number of realistic and distinct failure scenarios.
Better performance of fuzzing compared to the search-based approach can be explained by the formulation of the optimization objective, \ie{} the minimization of requirement satisfaction probability.
This objective tends to drive the search-based approach toward simpler models (\eg{} with fewer transitions), concentrating exploration near global optima in the parameter space.
Such models are likely to violate the specified requirements, yet they typically lead to clinically unrealistic behaviors. 
This plausibly explains the substantial discrepancy between the total number of detected failures and those deemed clinically realistic.

\begin{rqbox}[RQ2 summary.]
Fuzzing consistently outperforms both the search-based approach and random search, identifying the highest proportion of realistic failure scenarios (median values between $35$-$60\%$ across requirements) and producing more clusters (median values $18-19$).
The other methods yield substantially fewer realistic failures (around $10\%$) and less clusters (median values between $10$-$16$).
\end{rqbox}

\subsubsection{Cost of failure detection (RQ3)}
We use the same setup as in RQ2, but evaluate cost in terms of wall-clock execution time.
We record the execution time of both model space analysis and clustering for each method under comparison.

\begin{figure}[tb]
\centering
    \begin{subfigure}{0.5\columnwidth}
        \centering
        \includegraphics[width=\linewidth]{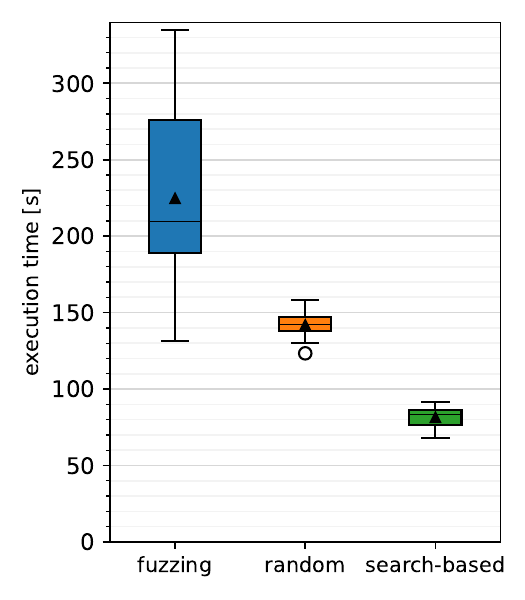}
        \caption{Wall-clock time.} \label{fig:times-boxplot-tot}
    \end{subfigure}%
    \begin{subfigure}{0.5\columnwidth}
        \centering
        \includegraphics[width=\linewidth]{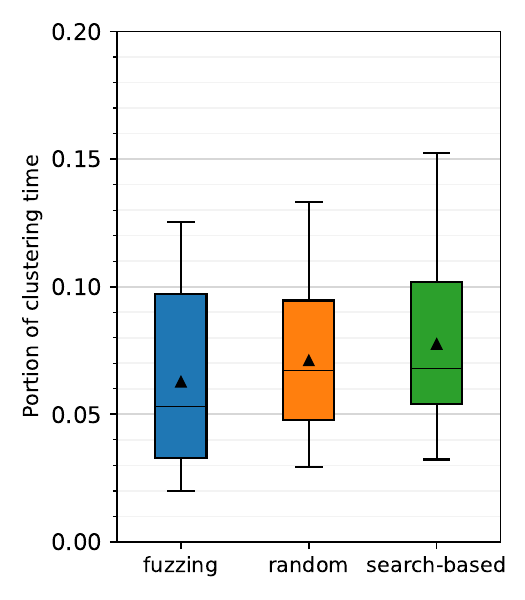}
        \caption{Partition of steps.} \label{fig:times-boxplot-frac}
    \end{subfigure}
    \caption{Cost of exploration techniques in execution time (RQ3).}
    \label{fig:times-boxplot}
\end{figure}

\vspace{.3em}\noindent\textbf{Results}.
Figure~\ref{fig:times-boxplot-tot} shows the distribution of execution time across $20$ repetitions.
The cost of fuzzing is generally higher and exhibits a higher variance (ranging from $2$ to $6$ minutes) compared to the other methods.
The search-based approach is the most time-efficient, typically finishing just over $1.5$ minutes, while random search averages around $2.5$ minutes.
Such efficiency can be attributed to its optimization objective, which explicitly minimizes requirement satisfaction probability.
As anticipated above (see RQ2), this objective steers the search toward simpler yet unrealistic models.
Such models are likely to violate the requirements and are usually faster to evaluate.

Overall, most of the execution time is spent performing \ac{SMC} with \uppaal{}, while the time overhead for mutant generation is generally negligible.
The clustering activity also represents a small fraction of the total time, usually around $5$-$10\%$, as shown in \fref{times-boxplot-frac}.

\begin{rqbox}[RQ3 summary.]
The search-based approach is the most time-efficient method, completing significantly faster than both fuzzing and random search, largely due to its optimization toward simpler models that are quicker to evaluate.
Execution time is dominated by \ac{SMC}, with mutant generation and clustering contributing only marginal overhead.
\end{rqbox}

\subsubsection{Effectiveness of stabilization strategies (RQ4)}
We assess the effectiveness of the proposed stabilization strategies by analyzing the patient's physiological state after the application of each strategy and comparing it against the outcome achieved by a human physician under the same clinical conditions.

To apply the strategies, we synthesized an automated controller using \uppaal{} \stratego~\cite{david2014time}.
To ensure statistical robustness, we repeat the strategy synthesis process $20$ times, verifying that the resulting strategies lead to consistent and reproducible outcomes.
We then execute $8$ simulated clinical scenarios in BREATHE, reproducing the health complications previously presented to the physician.
In each scenario, BREATHE's mechanical ventilator is automatically controlled according to the synthesized strategy, with patient states periodically matched against the mapping table to trigger the corresponding control actions.

At the end of a fixed observation period, we analyze the patient's physiological state by counting the number of monitored patient vital signals $v_i \in V$ that lie within their predefined safe operating ranges $R_i$, and comparing this count against the human baseline.
We consider a strategy as \emph{successful} if it stabilizes at least as many vital signals as the human physician.
In addition, we analyze the behavior of $\param{TV}$, the variable governed by flow conditions in the \ac{SHA} models.
For each scenario, we record the $\param{TV}$ value from BREATHE at the end of the observation period under the synthesized strategy and compare it against the simulated patient's nominal healthy value (equal to $400$\,ml for all scenarios); the physician's outcomes are evaluated using the same criterion.

\vspace{.3em}\noindent\textbf{Results}.
Overall, the application of the strategies synthesised by \nameExt{} is successful in $7$ out of $8$ scenarios.
\tref{stabilization} reports the deviation $\Delta$ of the observed $\param{TV}$ values from the nominal reference, expressed as both relative and absolute errors across all $8$ scenarios, with average values summarized in the final row.
On average, the automated controller maintains $\param{TV}$ approximately $20\%$ closer to the corresponding nominal value compared to the human physician.

\begin{table}[t]
    \centering
    \caption{Deviation compared to nominal $\param{TV}=400$\,ml (RQ4).  The bottom row shows average values.}
    \label{tab:stabilization}
    \begin{tabular}{c|c||c|c}
        \toprule
        \multicolumn{2}{c|}{Human physician} & \multicolumn{2}{|c}{Controller} \\
        \midrule
        absolute $\Delta$ [ml] & relative $\Delta$ [\%] & absolute $\Delta$ [ml] & relative $\Delta$ [\%] \\
        \midrule
        201 &  50.2 &  39 &   9.7 \\
        282 &  70.5 &  39 &   9.7 \\
        237 &  59.2 & 224 &  56.0 \\
        197 &  49.2 & 197 &  49.2 \\
        400 & 100.0 & 400 & 100.0 \\
        216 &  54.0 &  39 &   9.7 \\
        400 & 100.0 & 400 & 100.0 \\
        124 &  31.0 &  36 &   9.0 \\
        \midrule
        257.1 & 64.2 & 171.7 & 42.9 \\
        \bottomrule
    \end{tabular}
\end{table}

\begin{rqbox}[RQ4 summary.]
The stabilization strategies synthesized by \nameExt{} are successful in most cases, matching or outperforming the human physician.
On average, the automated controller maintains $\param{TV}$ approximately $20\%$ closer to its nominal healthy value, indicating comparable, and often improved, physiological stabilization relative to human-driven interventions.
\end{rqbox}

\section{Discussion}
\label{sec:discussion}

\subsection{Threats to Validity}

Threats to external validity arise from the use of a single evaluation subject.
We mitigate this threat by considering a wide variety of simulated clinical scenarios covering multiple health complications with varying severity levels, all validated by a physician with four years of working experience in intensive care units.
All experiments rely on synthetic data generated by the \textsc{Pulse} physiological engine, which is a high-fidelity simulator widely used for medical training and research.
The learning and validation phases involve a single physician, which may limit the generalizability of clinical judgments and realism criteria used to detect plausible failures.

To mitigate internal validity threats, all techniques are evaluated under identical computational budgets, identical mutation operators, and identical sets of requirements.
Each stochastic method is executed multiple times, and statistical testing is applied to reduce the risk of obtaining results by chance~\cite{ArcuriICSE2011}.

As reported in Section~\ref{sec:exp}, the search-based approach is optimized toward minimizing requirement satisfaction probability, which may bias it toward simpler and faster-to-evaluate models.
While this behavior is intrinsic to the optimization objective and intentionally analyzed, it may partially explain performance differences and must be considered when interpreting effectiveness and cost results.
Similarly, clustering-based diversity assessment depends on the chosen distance metric and clustering parameters, which may influence absolute cluster counts, although relative comparisons remain consistent across techniques.

Concerning the effectiveness of the strategies synthesized by \nameExt, we measured the number of stabilized vital signs and the deviation of $\param{TV}$ from its nominal healthy value.
This metric captures short-term physiological stabilization but does not account for possible longer-term outcomes (or secondary clinical objectives).
As a consequence, reported results should be interpreted as indicators of stabilization capability rather than comprehensive clinical optimality.
Furthermore, we assume that the actions selected by the synthesized controller are always applied to the patient.
This assumption models a deployment scenario in which the physician fully accepts the recommendations of a clinical decision support system, deliberately excluding cases of partial compliance or override.
Under this assumption, the reported results can be interpreted as a lower bound stabilization capability. 
Real-world effectiveness will depend on human-\ac{DT} interaction dynamics, which are outside the scope of this study.

We mitigate conclusion validity threats by repeating experiments across multiple runs and scenarios, applying non-parametric statistical tests where applicable, and reporting effect sizes alongside statistical significance.

\subsection{Impact}
This work contributes to the SAFEST project\footnote{\url{https://safest-prin.github.io}} by advancing trustworthy \ac{DT}-based methodologies for medical \acp{CPS} under uncertainty and heterogeneity.
By integrating probabilistic modeling, automated exploration of critical contingencies, and strategy synthesis within a closed-loop \ac{DT} architecture, the methodology increases confidence in \ac{DT}-driven decision support despite incomplete data and modeling approximations.
The explicit treatment of uncertainty and the modular representation of actors support sound dependability analysis while managing the complexity of heterogeneous \ac{DT} components.

The work also highlights two complementary needs central to SAFEST.
First, effective \ac{DT}-based analyses rely on reliable physiology simulators \cite{colombo2025breathe, bombarda2025simspire} to generate realistic data and stress-test rare critical conditions.
Second, automating the extraction of medical procedures and guidelines into machine-readable representations reduces ambiguity in practitioner behavior.
Approaches such as MARACTUS \cite{guindani2026agentic} address this need by transforming unstructured clinical documents into analyzable action models for integration into model-driven pipelines and \acp{CDSS}.
Together, these directions strengthen the SAFEST vision of dependable \ac{DT}-based healthcare solutions.

\section{Related Work}
\label{sec:rw}
{A review of the existing literature reveals that no prior study directly mirrors the approach proposed in this paper.}
A \ac{DT}-oriented architectural solution to formalizing the \ac{PDP} triad is introduced by Bersani et al.~\cite{bersani2022vision, bersani2022dttrust}.
Their approach emphasizes trust, defined as alignment between physical systems and \acp{DT}, and includes mechanisms for validating performance under human and environmental uncertainty~\cite{bersani2022vision}, a high-level reference architecture, and reliability estimation tools~\cite{bersani2022dttrust}.
Nonetheless, their work remains conceptual and lacks empirical evaluation.

The \ac{DT} concept originally introduced by Grieves~\cite{grieves2005newparadigm} has matured substantially, especially in manufacturing contexts.
Abanda et al.~\cite{abanda2025industry} survey ten \ac{DT} definitions from 2012 to 2022, illustrating this evolution.
With increasing human involvement in industrial processes, \acp{DT} began to model human aspects, leading to the development of \acp{HDT}.
Gaffinet et al.~\cite{gaffinet2025survey} review ten surveys on \acp{HDT}, highlighting their expansion across domains.
Miller et al.~\cite{miller2022humandt} define \ac{HDT} as a digital representation of a human combining mechanistic and statistical models, covering cognition, personality, and behavior. {Latif et al.~\cite{latif2020dtmanufacturing} provide a domain-specific example of human-performed process modeling.}
Lin et al.~\cite{lin2024hdt} survey \ac{HDT} technologies and propose a general architecture, focusing on modeling human activity and social behavior, but omitting task-based workflows such as medical procedures, which our model explicitly incorporates.

Asad et al.~\cite{asad2023hdt} examine Human-Centric \acp{DT}, noting that while ergonomics and health are addressed, cognitive processes and decision-making are generally overlooked.
Wang et al.~\cite{wang2024hdt} similarly highlight the overemphasis on physical assets at the expense of human modeling, proposing a layered \ac{HDT} architecture within which our \ac{SHA}-based patient model fits.
Our approach, however, introduces \ac{SMC}-based analysis, adding capabilities absent from their taxonomy.
Li et al.~\cite{li2024dhm} present a \ac{DT}-based framework for ergonomic analysis, focusing on body movement and cognition.
Their ``Digital Engine'' includes simulation and analysis components, mapping closely to our \ac{SHA}-based models and \ac{SMC}-driven dependability evaluation.
{Lauer-Schmaltz et al.~\cite{lauerSchmaltz2022hdt} review \acp{HDT} for behavioral therapy and rehabilitation, noting limited treatment of motivational and cognitive mechanisms.
Our work goes further by explicitly capturing medical knowledge and procedural workflows.}
Sahal et al.~\cite{sahal2022personaldt} propose a Personal \ac{DT} integrating blockchain and AI to support decision-making, though at a high level.
They identify healthcare requirements such as clinical decision support and analysis.
Our system can be seen as a concrete realization of these goals.

Katsoulakis et al.~\cite{katsoulakis2024dt} survey healthcare-focused \acp{DT}, primarily addressing patient care and organ-specific modeling for personalized medicine.
One application area, namely hospital management and care coordination, includes hospital workflows and staff modeling, aligning with our approach.
Yet, none of the surveyed works incorporate automata-based models or testing-inspired analysis, highlighting our contribution.
{Chen et al.~\cite{chen2024network} review communication technologies for \acp{HDT} in personalized healthcare, providing low-level networking insights not covered here.
Likewise, Papachristou et al.~\cite{papachristou2024precision} review \acp{DT} for precision medicine, covering diagnosis, surgery, and therapy, but do not address task-oriented physician modeling as we do.}

Roopa and Venugopal~\cite{roopa2025dt} focus on \acp{DT} in Cyber-Physical Healthcare Systems (\ac{DT}-CPHS), identifying operational efficiency as a key requirement.
Our system contributes by detecting critical scenarios in intensive care caused by physician misbehavior, enhancing care delivery.
Croatti et al.~\cite{croatti2020agents} integrate \acp{DT} with Multi-Agent Systems for smart healthcare, modeling heterogeneous entities and supporting trauma care, but their \ac{DT} primarily aids documentation and procedure tracking without advanced analytics.
Mariani et al.~\cite{mariani2023agents} extend this work conceptually, discussing simulation and prediction, yet without concrete implementation.
While aligned with representing the \ac{PDP} triad through \acp{DT}, our approach delivers a more operational and grounded realization.

Mohamed et al.~\cite{mohamed2023dt} propose a \ac{DT} framework for healthcare systems engineering, including patients, staff, and facilities, but model physicians and processes more as administrative units than as knowledge-driven clinical agents.
Our approach captures both behavioral dynamics and medical expertise of physicians within the \ac{DT}.
{Elayan et al.~\cite{elayan2021dt-iot} and Rahim et al.~\cite{rahim2024dt} both implement ECG classifiers in \ac{DT} frameworks.
Elayan et al. focus on real-time machine learning for heart disease diagnosis using IoT-integrated \acp{DT}, with healthcare improvements treated as secondary.
Rahim et al. present a Smart Medical \ac{CPS} for hospitals, integrating patients, devices, and staff.
Their system supports arrhythmia diagnosis via ECG classification and provides monitoring tools for healthcare professionals.
While both works offer predictive functionality, our approach uniquely integrates formal task modeling and automated analysis to identify critical care scenarios.}

\section{Conclusion}
\label{sec:concl}

\nameExt{} is an extended reliability-engineering methodology for the dependability analysis and attainment of critical \ac{PDP} systems in healthcare, operating within a closed-loop \ac{DT} paradigm.
By combining data-driven learning of patient dynamics with automata-based modeling, statistical model checking, model-space exploration, and diversity analysis, \nameExt{} can systematically identify, classify, and mitigate clinically plausible failure scenarios.
Fuzzing-based model-space exploration outperforms baseline approaches, detecting $35$–$60\%$ realistic failure scenarios and producing richer clusters of critical contingencies (median $18$–$19$ versus $10$–$16$).
Synthesized strategies stabilize patient vitals at least as effectively as human decision-making in $87.5\%$ of evaluated cases, keeping relevant metrics closer to nominal healthy values.

Scaling \nameExt{} to more complex clinical settings and enriching human-in-the-loop models with cognitive and team-level factors are future works.
Additional directions include improving robustness under partial observability, tighter integration with high-fidelity physiology simulators, and further automation of clinical procedure extraction.

\ifanonymous{}
\else{
\section*{Acknowledgments}
This work was partially supported by the PRIN project SAFEST (award No: 20224AJBLJ) funded by the Italian Ministry of Education, Universities, and Research (MIUR).
}
\fi

\bibliographystyle{plain}
\bibliography{biblio}

\end{document}